\documentclass[review,a4paper]{article}
\usepackage{amsmath,amssymb,amsfonts}
\usepackage[vlined, ruled, linesnumbered]{algorithm2e}
\usepackage[tight]{subfigure}
\usepackage[T1]{fontenc}
\usepackage[utf8]{inputenc}
\usepackage{tikz}
\usepackage{stmaryrd}

\newtheorem{theorem}{Theorem}

\newtheorem{corollary}{Corollary}

\DeclareMathOperator*{\dist}{dist}
\DeclareMathOperator*{\bw}{bw}
\DeclareMathOperator*{\tw}{tw}

\def\algname{\texttt{Determine-Hom} }
\def\0{{\bar 0}}
\def\a{{\bf a}}
\def\b{{\bf b}}
\def\p{{\bf p}}
\def\vv{{\mathbf v}}

\begin{document}

\title{Exact Algorithm for Graph Homomorphism\\ and Locally Injective Graph Homomorphism}

\author{
Pawe{\l} Rz{\k a}{\. z}ewski  \\
\texttt{p.rzazewski@mini.pw.edu.pl} \\ \\
Warsaw University of Technology \\
Koszykowa 75 , 00-662 Warsaw, Poland
}
\date{ }
\maketitle

\begin{abstract}
For graphs $G$ and $H$, a homomorphism from $G$ to $H$ is a function $\varphi \colon V(G) \to V(H)$, which maps vertices adjacent in $G$ to adjacent vertices of $H$. A homomorphism is locally injective if no two vertices with a common neighbor are mapped to a single vertex in $H$. Many cases of graph homomorphism and locally injective graph homomorphism are NP-complete, so there is little hope to design polynomial-time algorithms for them. In this paper we present an algorithm for graph homomorphism and locally injective  homomorphism working in time $\mathcal{O}^*((b + 2)^{|V(G)|})$, where $b$ is the bandwidth of the complement of $H$.
\end{abstract}

\section{Introduction}
Graphs homomorphism problem (or $H$-coloring, as it is sometimes called) is a natural generalization of a well-known graph coloring problem. For graphs $G$ and $H$ we say that $\varphi \colon V(G) \to V(H)$ is a homomorphism from $G$ to $H$ is $\varphi(v)\varphi(u) \in E(H)$ for any $uv \in V(G)$. In other words, a homomorphism is an edge-preserving mapping from $V(G)$ to $V(H)$. Thus $k$-coloring problem is equivalent to the problem of finding a homomorphism to the complete graph $K_k$. 
We refer the reader to the monography by Hell and Ne\v set\v ril \cite{HN} for more information about graph homomorphisms.

Some special cases of graph homomorphism, namely {\em locally constrained graph homomorphisms} have also received a considerable attention. We say that the homomorphism $\varphi$ is {\em locally injective} ({\em locally surjective}; {\em locally bijective}) if the neighborhood of $v \in V(G)$ is mapped injectively (resp.: surjectively; bijectively) to the neighborhood of $\varphi(v)$. 
See the survey by Fiala and Kratochv\'il \cite{FK2} for more information about locally constrained homomorphisms.

Homomorphisms and locally constrained homomorphisms generalize and unify many known problems in graph theory, e.g. colorings,  graph covers, role assignments etc.
For example $(m,k)$-coloring (see Zhu \cite{Zhu}), i.e. the problem of finding an assignment $f \colon V(G) \to \{0,1,..,m-1\}$, such that $k \leq |f(v) - f(u)| \leq m-k$ whenever $vu \in E(G)$, is equivalent to finding a homomorphism to $\overline C^{k-1}_m$, a complement of a $(k-1)$'th power of an $m$-cycle\makeatletter{\renewcommand*{\@makefnmark}{$^\dag$}\footnote{We consider $C^0_m$ to be a graph with $m$ vertices and no edges.}.

Another example is the so-called $H(2,1)$-labeling problem (where $H$ is a graph), in which vertices adjacent in $G$ are mapped onto distinct, nonadjacent vertices of $H$, and vertices, which have a common neighbor in $G$, are mapped onto distinct distinct vertices of $H$. Fiala and Kratochv\'il \cite{FK} showed a close relation between locally injective homomorphisms and $H(2,1)$-labelings: an $H(2,1)$-labeling of $G$ is exactly a locally injective homomorphism from $G$ to $\overline H$, where $\overline H$ denotes the complement of $H$.

A well-known $L(2,1)$-labeling problem (see Griggs, Yeh \cite{GY}) can be seen as a problem of finding the minimum $k$ such that the input graph admits an $H(2,1)$-labeling for $H$ being a path with $k+1$ vertices.
Another interesting case of $H(2,1)$-labeling is a circular $L(2,1)$-labeling, sometimes denoted by $L_c(2,1)$-labeling (see Liu,Zhu \cite{LZ}). It is equivalent to finding the minimum $k$, for which the input graph $G$ has an $H(2,1)$-labeling for $H$ being a cycle with $k+1$ vertices.

Graph homomorphisms are also interesting from the computational point of view. In their celebrated theorem, Hell and Ne\v set\v ril \cite{HN2} showed that determining if $G$ has a homomorphism to $H$ is polynomial if $H$ is bipartite and NP-complete otherwise.
For a locally surjective homomorphism Fiala and Paulusma~\cite{FP} showed that determining the existence of a locally surjective homomorphism from $G$ to a connected graph $H$ is polynomial if $H$ has at most 2 vertices and NP-complete otherwise. They also showed a full dichotomy for the case when $H$ is disconnected, but the description of polynomial cases is more complicated.
There is no similar characterization for the case of locally injective homomorphisms and locally bijective homomorphisms, but still we can find some partial results (see for example \cite{FK,FK3,FKP} for some results on locally injective homomorphisms and \cite{AFS,KPT} for locally bijective homomorphisms).

As many cases of graph homomorphism and locally constrained graph homomorphism are NP-complete, there is little hope to obtain polynomial algorithms for them. Therefore a natural approach is to design exponential algorithms with the basis of the exponential factor in a complexity bound expressed  as a function of some invariant of $H$. Fomin {\em et al.} \cite{FHK} presented the algorithm for graph homomorphism from $G$ to $H$ working in time $\mathcal{O}^*((2\tw(H)+1)^n)$\makeatletter{\renewcommand*{\@makefnmark}{$^\ddag$}\footnote{In the $\mathcal{O}^*$ notation we suppress polynomial factors.}\makeatother}, where $\tw(H)$ denotes a {\em treewidth} of the graph $H$ (see Diestel's book \cite{Diestel} for some information about treewidth of graphs) and $n$ is the number of vertices of $G$.
For a locally injective homomorphisms, Havet {\em et al.} \cite{HKKKL} presented an algorithm working in time $\mathcal{O}^*((\Delta(H)-1)^n)$. To our best knowledge there are no similar results for a locally surjective and a locally bijective graph homomorphism problem.

In this paper we show how to adapt the algorithm for $L(2,1)$-labeling by Junosza-Szaniawski {\em at al.} \cite{TAMC} to solve graph homomorphism and locally injective graph homomorphism problems.
We have already mentioned that graph homomorphism to a complete graph is equivalent to graph coloring problem and therefore can be solved in time $\mathcal{O}^*(2^n)$, using the algorithm by Bj\"orklund {\em et al.} \cite{BHK}. Finding a locally injective homomorphism to a complete graph can also be reduced to the classical graph coloring, since it is equivalent to coloring a square of the graph. Therefore in this paper we shall focus on the case when $H$ is not a complete graph.
The main result of this paper is the following theorem.

\begin{theorem} \label{thm-hom}
The existence of a homomorphism from $G$ to $H$ can be decided in time $\mathcal{O}^* \left((\bw(\overline H)+2)^n \right )$, where $n$ is the number of vertices of $G$ and $\bw(\overline H)$ is the bandwidth of the complement of $H$.
\end{theorem}

After a small adaptation in the algorithm for the graph homomorphism problem, we obtain the algorithm for the locally injective homomorphism problem, working within the same time bound.

\begin{theorem} \label{thm-lih}
The existence of a locally injective homomorphism from $G$ to $H$ can be decided in time $\mathcal{O}^* \left((\bw(\overline H)+2)^n \right )$, where $n$ is the number of vertices of $G$ and $\bw(\overline H)$ is the bandwidth of the complement of $H$.
\end{theorem}

\section{Preliminaries}
In his paper we consider simple graphs without loops and multiple edges.
For a graph $G=(V,E)$, let $\overline{G}$ denote its complement, i.e. $\overline{G} = (V, \binom{V}{2} \setminus E)$.

For graphs $G$ and $H$, a {\em homomorphism} from $G$ to $H$ is a mapping $\varphi \colon V(G) \to V(H)$ such that $\varphi(v)\varphi(w) \in E(H)$ whenever $vw \in E(G)$. 
If $\varphi$ is a homomorphism from $G$ to $H$, we will write $\varphi \colon G \to H$ shortly.
A homomorphism is {\em partial} if we allow that some vertices of $G$ are not mapped to any vertices of $H$.

A homomorphism $\varphi$ from $G$ to $H$ is {\em locally injective} if a neighborhood of any vertex $v \in V(G)$ is mapped injectively into the neighborhood of $\varphi(v)$. In other words, no two vertices from $G$ with a common neighbor are mapped to the same vertex of $H$.

An $H(2,1)$ labeling of $G$ is a mapping $\psi \colon V(G) \to V(H)$, such that (1) $\dist_H(\psi(v),\psi(w)) \geq 2$ if $\dist_G(v,w)=1$ and (2) $\dist_H(\psi(v),\psi(w)) \geq 1$ if $\dist_G(v,w)=2$.
According to Fiala and Kratochv\'il \cite{FK}, an $H(2,1)$-labeling of $G$ is exactly a locally injective homomorphism from $G$ to $\overline H$.

Let $G$ be a graph and let $L = v_1v_2\ldots v_n$ be some ordering of its vertices.
A {\em bandwidth} of a graph $G$, denoted by $\bw(G)$, is the minimum over all orderings $L$ of the value $\max \{|i-j| \colon v_iv_j \in E(G)\}$. Informally speaking, we want to place the vertices of $G$ on integer points of a number line in such a way, that the longest edge is as short as possible.

For $\ell \in \mathbb{N}$ let $[\ell]$ denote the set $\{0,1,2,..,\ell\}$. Moreover,  let $\llbracket \ell \rrbracket$ denote the set $[\ell] \cup \{\bar 0\}$, where $\bar 0$ is a special symbol, whose meaning will be clarified later.

For a set of vectors $A \subseteq \Sigma^n$ and a symbol $x \in \Sigma$ let $A_x$ denote the set $\{\vv \in \Sigma^{n-1} \colon x\vv \in A\}$ where $x\vv$ denotes concatenating $x$ and $\vv$.

\section{Exact Algorithm for Graph Homomorphism}

Let us consider a problem of deciding if a graph $G$ has a homorphism to $H$.
Let $V(G) = \{v_1,v_2,..,v_n\}$ and $V(H) = \{h_1,h_2,..,h_m\}$. The ordering of vertices in $G$ is arbitrary. The vertices of $H$ are arranged in the order corresponding to the bandwidth of $\overline{H}$, i.e. in such a way that the value $\max( \{ |i-j| \colon h_ih_j \notin E(H)\}$ is minimum possible (recall that this minimum value is equal to $\bw(\overline H)$). Let $\beta = \bw(\overline H)+1$ and let $H_k = H[\{h_1,h_2,\ldots,h_k\}]$ for any $k \in \{1,2,\ldots,m\}$.

In this section we prove Theorem \ref{thm-hom} by presenting an algorithm from determining the existence of the homomorphism $G \to H$, working in time $\mathcal{O}^* \left((\bw(\overline H)+2)^n \right )$.
We shall proceed in a way similar to the algorithm by Junosza-Szaniawski {\em et al.}~\cite{TAMC}. We will use dynamic programming and try to extend partial homomorphisms from $G$ to $H_k$ to partial homomorphisms from $G$ to $H_{k+1}$.

Let $P$ be a set of characteristic vectors of independent sets in $G$. In other words, $\p \in P$ if and only if there exists an independent set $X$ in $G$ such that $\p_i = 1$ iff $v_i \in X$ and $\p_i = 0$ otherwise.

For every $k=1,..,m$ we introduce a set $T[k] \subseteq [\beta]^n$ such that $\a \in T[k]$ if and only if there exists a partial homomorphism $\varphi \colon G \to H_k$, satisfying the following condition:

\begin{center}
$\a_i = \begin{cases}
0 & \text{if $v_i$ is not mapped},\\
1 & \text{if $\varphi(v_i) = h_\ell$ with $\ell \leq k-\beta+1$}, \\
x \in \{2,..,\beta\} & \text{if $\varphi(v_i)=h_\ell$ with $\ell=k-\beta+x$}.\\
\end{cases}$
\end{center}

Moreover, let us define $T[0] := \{0^n\}$.
Note that vectors $\a$ with no $0$'s correspond to homomorphisms $G \to H_k$. Therefore there exists a homomorphism $G \to H$ if and only if $T[m] \cap \{1,2,\ldots,\beta\}^n \neq \emptyset$. 

To compute sets $T[k]$ (for $k \in \{1,\ldots,m\}$) we shall introduce two operations.
Let $k \in \{1,\ldots,m-1\}$ be fixed and assume we have computed the set $T[k]$. Let $\a$ be a vector from $T[k]$.
Let $\overline{\a} \in \llbracket \beta \rrbracket^n$ be defined as follows:
$$\overline \a_i = \begin{cases}
0 & \text{if $\a_i=0$ and there is no $v_j\in N_G(v_i)$}\\
 & \text{with $\a_j \geq 2$ and $h_{k-\beta+\a_j} \notin N_H(h_{k+1})$},\\
\0 & \text{if $\a_i=0$ and there exists $v_j\in N_G(v_i)$}\\
 & \text{with $\a_j \geq 2$ and $h_{k-\beta+\a_j} \notin N_H(h_{k+1})$},\\
x \in \{1,..,\beta\} & \text{if $\a_i = x$}.
\end{cases}$$

Notice that for any $\a \in T[k]$ (for $k \leq m-1$) and partial homomorphism $\varphi$ corresponding to $\a$, we have $\overline \a_i = 0$ if and only if $\varphi$ can be extended by mapping $v_i$ to $h_{k+1}$. Since the vertices of $H$ are arranged according to $\bw(\overline H)$, all non-neighbors of $h_{k+1}$ are in $\{h_{k-\bw(\overline H)+1},h_{k-\bw(\overline H)+2},\ldots,h_k\}$. This justifies the unification of the sets of vertices mapped to $h_1,h_2\ldots,h_{k-\bw(\overline H)}$ in our representation of partial homomorphisms.
Let $\overline{T}[k]$ denote the set $\{\overline {\a} \colon \a \in T[k]\}$. Note that 
computing $\overline T[k]$ from $T[k]$ takes time $\mathcal{O} (|T[k]| \cdot n^2)$, since in each vector $\a \in T[k]$ we just have to check all pairs of vertices $v_i,v_j$ such that $v_iv_j \in E(G)$ and $\a_i=0$. Moreover observe that $|\overline{T}[k]| = |T[k]|$ for every $k$.

Now let us define the partial function $\oplus  \colon \llbracket \beta \rrbracket \times\{0,1\}\to [\beta]$ as follows:
\begin{center}
$
x \oplus y = \left \{
\begin{array}{l l}
0 & \text{if $x \in \{0,\0\}$ and $y=0$}, \\
1 & \text{if $x \in \{1,2\}$ and $y=0$}, \\
x-1 & \text{if $x \in \{3,4,..,\beta\}$ and $y=0$}, \\
\beta & \text{if $x =0$ and $y=1$}, \\
\text{undefined} & \text{otherwise.}
\end{array}
\right .
$
\end{center}
We generalize this operation to vectors coordinate-wise ($x_1x_2.. x_n\oplus y_1y_2.. y_n$ is $(x_1\oplus y_1)..(x_n\oplus y_n)$  if $x_i\oplus y_i$ is defined for all $i\in\{1,..,m\}$ or is undefined otherwise);
and sets of vectors: $A\oplus B= \{\,\a\oplus\b \colon \a\in A,\ \b\in B,\ \a\oplus\b \text{ is
defined}\,\}$.

Observe that for $\a \in T[k]$ and $\p \in P$, computing $\overline \a\oplus \p$ corresponds to
extending a partial homomorphism by
mapping all the vertices from the set encoded by $\p$ to vertex $h_{k+1}$.
Now we are ready to present the algorithm.

\begin{algorithm}[H]
\label{alg-hom}
\caption {\algname$(G,H)$}
$P \gets $ a set of characteristic vectors of independent sets of $G$\\
$T[0] \gets \{0^n\}$\\
\For {$k \gets 1$ \KwTo $m$}
{
compute $\overline{T}[k-1]$ from $T[k-1]$\\
$T[k] \gets \overline T[k-1] \oplus P$\\
}
\lIf {$T[m] \cap \{1,2,\ldots,\beta\}^n \neq \emptyset$} {\Return {\sc Yes}}\\
\lElse {\Return {\sc No}}
\end{algorithm}

To prove the correctness, it is enough to show that $T[k] = \overline T[k-1] \oplus P$ for any $k=1,2,\ldots,m$.  Let $\a \in T[k]$ and $\varphi \colon G \to H_k$ be a partial homomorphism corresponding to $\a$. Let $\p$ be a characteristic vector of the set $X:=\varphi^{-1}(h_k)$. This set is clearly independent, so $\p \in P$. Let $\varphi' \colon G \to H_{k-1}$ be a partial homomorphism obtained from $\varphi$ by unmapping all vertices from $X$, and let $\a'$ be a vector in $T[k-1]$ corresponding to $\varphi'$. Let $v_i$ be a vertex from $X$. Every neighbor of $v_i$ has to be mapped to some neighbor of $h_k$ or be not mapped at all, so $\a'_i =0$. We observe that $\a = \overline \a' \oplus \p$ and therefore $\a \in \overline T[k-1] \oplus P $.

On the other hand, let $\a' \in T[k-1]$ (with a corresponding partial homomorphism $\varphi' \colon G \to H_{k-1}$) and $\p \in P$. Let us extend $\varphi'$ to partial homomorphism $\varphi \colon G \to H_{k}$ with every  vertex $v_i$ from the independent set corresponding to $\p$ mapped to vertex $h_k$. It is possible if and only if each neighbor of $v_i$ is not mapped or is mapped to a neighbor of $h_k$. In other words, we require that $\overline \a_i = 0$. Therefore $\overline \a' \oplus p \in T[k]$ with corresponding partial homomorphism $\varphi$.

We shall perform the computation of $\overline T[k-1] \oplus P$ recursively. We consider vectors starting with each element from $[\beta]$ separately.
\vskip -0.8cm
\begin{align*}
\overline T[k]\oplus P =&  \bigcup_{\substack{b \in \llbracket \beta \rrbracket, p \in \{0,1\}\\ \text{s.t. } b \oplus p  \text{ is defined}}}  (b \oplus p) (\overline T[k]_b \oplus P_p)\\
=  &  ~0 \left [ ( \overline{T}[k]_0 \cup \overline{T}[k]_{\overline{0}}) \oplus P_0 \right ]  
\cup  1 \left [ \left( \overline{T}[k]_1 \cup \overline{T}[k]_2\right) \oplus P_0 \right ]  \\
\cup  & ~\bigcup_{a \in \{2,..,\beta-1\}} a \left [ T[k]_{a+1} \oplus P_0 \right ] 
\cup  ~\beta \left [ T[k]_{0}  \oplus P_1 \right ].
\end{align*}

To compute $\oplus$ on two sets of vectors of length $n$, we have to compute $\oplus$ on $\beta+1$ pairs of sets of vectors of length $n-1$.
The size of $P$ is at most $n \cdot 2^n$ bits, the size of $T[k-1]$ is at most $n \cdot (\beta+1)^n$ bits. Recall that computing $\overline T[k-1]$ takes time at most $\mathcal{O} (|T[k-1]| \cdot n^2) = \mathcal{O} (n^3 \cdot (\beta+1)^n)$. 
Therefore time complexity is given by:
$F(n) =  \mathcal{O} \bigl( n \cdot 2^n + n^3 \cdot (\beta+1)^n + (\beta+1) \cdot F(n - 1)  \bigr).$
One can verify by induction that this recursion is satisfied by $F(n)= \mathcal{O} \bigl(n^3 \cdot 2^n +  n^3 \cdot (\beta+1)^n \bigr )$.
Recall that $\beta = \bw(\overline H)+1$ and therefore $\beta+1 =\bw(\overline H) +2 \geq 2$. Finally we obtain
$F(n) = \mathcal{O}^* \bigl((\bw (\overline H)+2)^n \bigr )$, which proves Theorem \ref{thm-hom}.

Recall that $(m,k)$-coloring is equivalent to a homomorphism to $\overline C^{k-1}_m$. Since a complement of $\overline C_m^{k-1}$ is $C_m^{k-1}$ and $\bw (C_m^{k-1}) = 2(k-1)$, we obtain the following.

\begin{corollary}
The $(m,k)$-coloring problem on a graph $G$ with $n$ vertices can be solved in time $\mathcal{O}^* \left((2k)^n \right )$.
\end{corollary}

\section{Locally Injective Homomorphism and $H(2,1)$-labeling}

In this section we prove Theorem \ref{thm-lih} by modifying Algorithm \algname to determine the existence of \textbf{a locally injective} homomorphism from $G$ to $H$.

We observe that the vertices of $G$ that can be mapped to a single vertex of $H$ (in a locally injective manner) must be in a distance at least 3 from each other. Therefore they form a 2-independent set, which is exactly an independent set in a square of the graph (see for example \cite{IPL} for more information about such sets).
Since it is the only additional requirement for locally injective homomorphisms, the only thing that has to be changed in Algorithm \algname is the initialization of the set $P$. Now it has to contain characteristic vectors of all 2-independent sets. This proves Theorem \ref{thm-lih}.

Recall that $H(2,1)$-labelings are exactly locally injective homomorphisms to $\overline H$. Since the complement of $\overline H$ is $H$, we obtain the following corollary.
\begin{corollary}
For any graphs $G$ and $H$ we can solve the $H(2,1)$-labeling problem in time $\mathcal{O}^*\left((\bw(H)+2)^n \right )$, where $n$ is the number of vertices of $G$.
\end{corollary}

Let us see how this bound works for the $L_c(2,1)$-labeling problem. Recall it is equivalent to finding the smallest $m$, such that the input graph admits a $C_m (2,1)$-labeling. We shall check the existence of such a labeling for $m=3,..,2n$ and stop when we find one. Since $\bw(C_m)=2$, we obtain the following.

\begin{corollary}
The $L_c(2,1)$-labeling problem on a graph $G$ with $n$ vertices can be solved in time $\mathcal{O}^* \left(4^n \right )$.
\end{corollary}

The bounds presented here can be slightly improved, using the methods presented in \cite{IPL} and \cite{TAMC}. However, it requires many technical calculations and the improvement gets smaller as $\bw(\overline H)$ grows.

\vskip 0.5cm
\noindent \textbf{Acknowledgement.} The author is sincerely grateful to Konstanty Junosza-Szaniawski for valuable discussion and advice.

\end{document}